\documentclass[a4paper,11pt]{article}
\usepackage{pos}
\usepackage{lineno}
\usepackage{hyperref}
\usepackage{caption}
\usepackage{subcaption}
\usepackage[numbers,sort&compress]{natbib}

\title{Study of the hadron gas phase using short-lived resonances with ALICE}

\author*[a]{Johanna L\"omker}

\onbehalf{on behalf of the ALICE collaboration}

\affiliation[a]{Nikhef,\\
  Science Park 105, 1098 XG Amsterdam, The Netherlands}

\affiliation[a]{GRASP, Utrecht University,\\
Princetonplein 1, 3584 CC Utrecht, The Netherlands}

\emailAdd{johanna.lomker@cern.ch}

\abstract{
Short-lived hadronic resonances are unique tools for studying the hadron-gas phase that is created in the late stages of relativistic heavy-ion collisions. Measurements of the yield ratios between resonances and the corresponding stable particles are sensitive to the competing rescattering and regeneration effects. These measurements in small collision systems, such as pp and p-Pb, are a powerful method to reveal a possible short-lived hadronic phase. In addition, resonance production in small systems is interesting to study the onset of strangeness enhancement, collective effects, and the hadron production mechanism. On this front, the $\phi$ meson is particularly relevant since its yield is sensitive to different production models: no effect is expected by strange number canonical suppression but its production is expected to be enhanced in the rope-hadronization scenario. 

In this presentation, recent measurements of hadronic resonances in different collision systems, going from pp to Pb-Pb collisions, are presented. These include transverse momentum spectra, yields, and yield ratios as a function of multiplicity. The presented results are discussed in the context of state-of-the-art phenomenological models of hadron production. The resonance yields measured in Pb-Pb collisions are used as an experimental input in a partial chemical equilibrium-based thermal model to constrain the kinetic freeze-out temperature. This is a novel procedure that is independent of assumptions on the flow velocity profile and the freeze-out hypersurface.
}

\FullConference{%
  EPS-HEP 2023 conference\\
  20-25th August 2023\\
  University of Hamburg, Hamburg, Germany
}

\begin{document}
\maketitle

\section{Introduction}
Short lived resonances can provide insights about the late stages of the evolution of heavy-ion collisions, known as hadronic-gas phase, due to their lifetime being similar to the decoupling time of the system. When the quark gluon plasma (QGP) expands and cools down the total particle yields are in principle determined at the chemical freeze-out, but it is assumed that (pseudo) elastic scatterings and regeneration effects of the decaying particles before the kinetic freeze-out counteract \cite{PhiK}. The (pseudo) elastic scattering causes a loss of correlation of the decay daughters from a given resonance and thus effectively decreases the particle yields. Possible regeneration effects, where two decay products regenerate into a given hadronic resonance, can increase the particle yields. Both effects would modify the spectral shapes of the measured particles depending on their time of production, duration of the hadronic-gas phase, the hadronic cross-section, and the lifetime of the resonance. Therefore the two competing effects are analysed through resonances with different lifetimes along with their ratios to the corresponding stable counterparts. The very short-lived $K^{*}(892)^{0}$ ($\tau \approx 4.1 fm/$\textit{c} \cite{pdg}) and $\Lambda(1520)$ ($\tau \approx 12.6 fm/$\textit{c}) are likely to decay early in the hadronic-gas phase. Thus the decay daughters of these resonances are assumed to be subjects of re-scattering (or regeneration) at low transverse momentum ($p_{T} \leq 3$ $\mathrm{GeV}/$\textit{c}), while the $\phi(1020)$ ($\tau \approx 46.3 fm/$\textit{c}) could be rather unaffected \cite{PhiK} \cite{Lambda}. The resonance yields and ratios measured in Pb-Pb are further compared to the proton-proton reference systems to understand if these effects arise due to the presence of the QGP, and therefore are exclusive for the evolution of heavy-ion collisions, or not. 

\section{Analysis methodology} 
The experimental measurement is performed with unique particle identification and tracking down to momenta of a few hundred $\mathrm{MeV}$ with the central barrel detectors of the ALICE experiment \cite{L12}. Vertexing is performed with the most central \textbf{I}nner-\textbf{T}racking-\textbf{S}ystem (ITS) which also provides the first measurement for the tracking, accompanied by the tracking information coming from the \textbf{T}ime \textbf{P}rojection \textbf{C}hamber (TPC). Apart from tracking the TPC is mainly used for the particle identification (PID) that is further matched to the information coming from the \textbf{T}ime-\textbf{O}f-\textbf{F}light (TOF). The V0 scintillator detectors, located close to the beam pipe, are employed for the triggering and multiplicity estimation.
The resonance yields are obtained through the invariant mass reconstruction with the help of the corresponding detected daughter particles in the decay channels $\Lambda(1520) \rightarrow pK^{-} (\Bar{p}K^{+}$) \cite{Lambda}, $K^{*}(892)^{0} \rightarrow \pi^{+} \pi^{-}$ and $\phi(1020) \rightarrow K^{+}K^{-}$ \cite{PhiK} . In both analyses a \textit{mixed-event-technique} is used to reject the combinatorial background. To extract the signal from the residual background, which is of similar shape as a Maxwell-Boltzmann distribution, a global fit with a convolution of a non-relativistic Breit-Wigner function and the Gaussian detector resolution is performed in each $p_{T}$ and centrality bin. The extracted raw signal yield is then corrected for the decay branching fraction, detector acceptance, reconstruction efficiency, track selection and PID efficiency with HIJING \cite{K53} MC simulations for the Pb-Pb analysis (where additional resonances are injected), PYTHIA Monash2013 \cite{K52, K64} for the pp analysis and GEANT3 \cite{L22} for the transport through the detector material. 

\section{Results and discussion}
\subsection{Resonance yields and mean transverse momenta}
\begin{figure}[h!]
     \centering
     \begin{subfigure}[l]{0.49\textwidth}
         \centering
        \includegraphics[width=\textwidth]{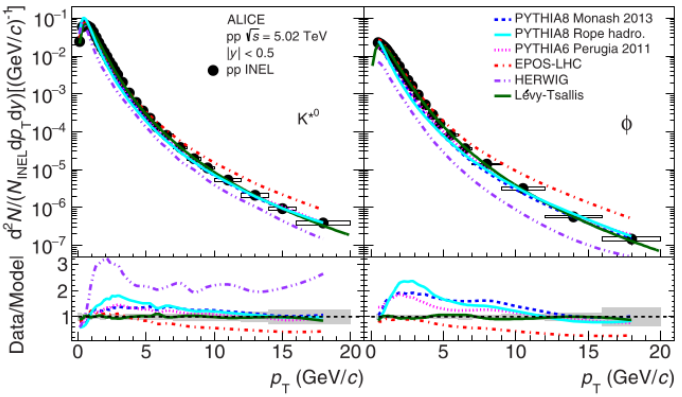}
    \end{subfigure}
    \hfill
    \begin{subfigure}[r]{0.49\textwidth}
         \centering
        \includegraphics[width=\textwidth]{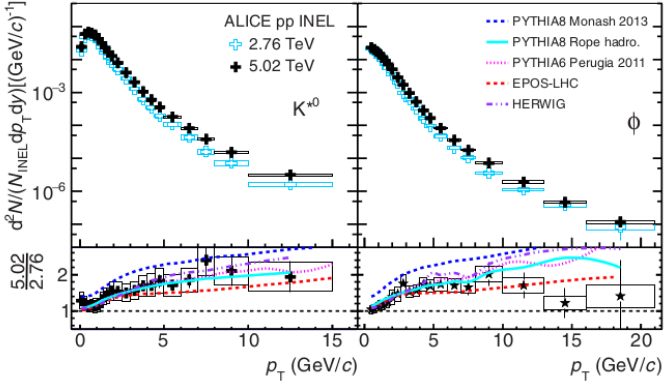}
    \end{subfigure}
    \caption{Comparisons for $K^{*0}$ and $\phi$ $p_{T}$-differential yields from pp-collisions (left; from \cite{PhiK}), for two different $\sqrt{s}$ (right; open markers correspond to the $2.76$ $\mathrm{TeV}$ results \cite{K21} and  the $5.02$ $\mathrm{TeV}$ results are from \cite{PhiK}) to model calculations from PYTHIA 6.4 (Perugia 2011 Tune) \cite{K62, K63}, PYTHIA 8.1 (Monash 2013 Tune), PYTHIA 8.2 (Rope hadronization) \cite{K65}, EPOS-LHC \cite{K66}, and HERWIG 7.1 \cite{K67}.}
    \label{fig:SpectraKPhi}
\end{figure}
The $p_{T}$-differential resonance yields for the $K^{*}(892)^{0}$ and $\phi(1020)$-mesons from pp-collisions are presented in Fig.\ref{fig:SpectraKPhi}. From the left figures, especially from the lower panels, it is apparent that none of the state-of-the-art models can describe any of the two resonance yields over the full $p_{T}$-range. The right figure shows the spectra for two different collision energies and indicates a subtle growth of the resonance yields with increasing energies. This increase of the yields for larger energies is, as can be seen from the lower panel, a $p_{T}$-dependent effect that saturates towards higher $p_{T}$. And even though none of the models could describe the spectral shapes well, the ratio between the two energies is well reproduced by all models except for PYTHIA Monash 2013 which overestimates the ratio.

\begin{figure}[h!]
     \centering
     \begin{subfigure}[l]{0.49\textwidth}
         \centering
        \includegraphics[width=\textwidth]{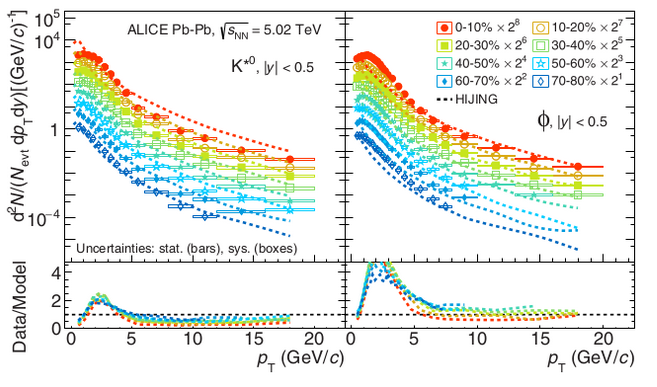}
    \end{subfigure}
    \hfill
    \begin{subfigure}[r]{0.49\textwidth}
         \centering
        \includegraphics[width=\textwidth]{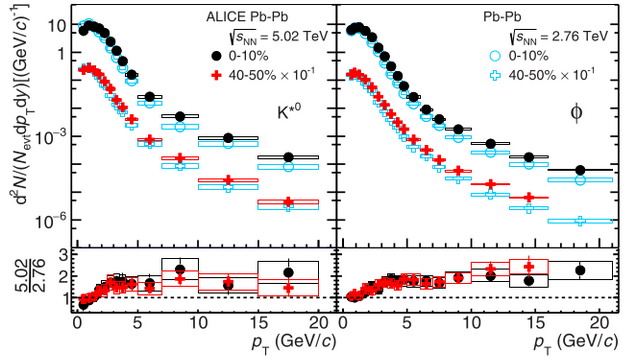}
    \end{subfigure}
    \caption{HIJING model comparisons for $K^{*0}$ and $\phi$ $p_{T}$-differential yields from Pb-Pb collisions (left), for two different $\sqrt{s_{NN}}$ and centrality classes (right) \cite{PhiK, K21, K18}.} 
    \label{fig:MultSpectraKPhi}
\end{figure}

As expected for Pb-Pb collisions, the resonance yields increase for more central collisions in the left plots of Fig.\ref{fig:MultSpectraKPhi}. Even though the overall evolution with centrality can be described by the HIJING model, the lower panels clearly point out that the model description fails to cover the full $p_{T}$-range for Pb-Pb collisions. The right plot of Fig.\ref{fig:MultSpectraKPhi} shows the resonance yields for two collision energies and, additionally, two different centrality ranges. The yields increase slightly for the higher $\sqrt{s_{NN}}$ while the energy ratio first grows for the central and mid-central collisions and then saturates towards larger $p_{T}$. The most important feature of this plot is that the $K^{*}(892)^{0}$ energy ratio around $p_{T} \leq 3$ $ GeV$/\textit{c} drops below unity, while the $\phi(1020)$ does not. This effect, at low $p_{T}$, is attributed to the re-scattering phenomena which are expected for the $K^{*}(892)^{0}$. The signal loss due to re-scattering is enhanced for the higher collision energies due to the denser system with higher pressure gradients produced with respect to the lower $\sqrt{s_{NN}}$.

\begin{figure}[h!]
     \centering
     \begin{subfigure}[l]{0.59\textwidth}
         \centering
        \includegraphics[width=\textwidth]{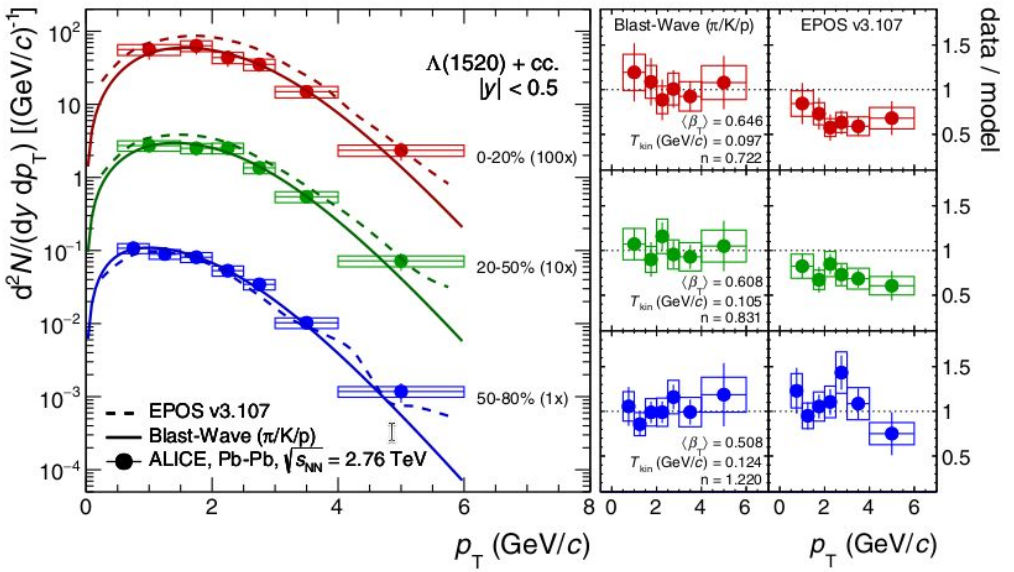}
    \end{subfigure}
    \hfill
    \begin{subfigure}[r]{0.33\textwidth}
         \centering
        \includegraphics[width=\textwidth]{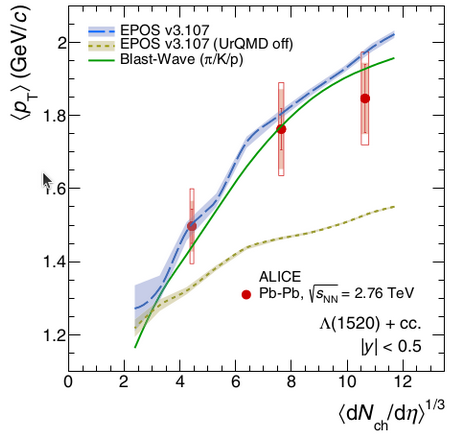}
    \end{subfigure}
    \caption{Blast-Wave and EPOS v3.107 (UrQMD) model comparisons for $\Lambda(1520)$ yields from Pb-Pb collisions as a function of $p_{T}$ (left) and $\langle p_{T} \rangle$ (right) as a function of system size ($\langle dN_{ch}/d\eta \rangle^{1/3}$) \cite{Lambda, L27, L17}.}
    \label{fig:MultSpectraLambda}
\end{figure}

While most models fail to describe the $K^{*}(892)^{0}$ and $\phi(1020)$ resonances in Fig.\ref{fig:MultSpectraKPhi} over the full $p_{T}$-range the Blast-Wave model \cite{L23, L24}, tuned to measured $\pi, K$ and $p$ yields, serves as a good description for all centrality classes of the $p_{T}$-differential $\Lambda(1520)$ resonance yields in Fig.\ref{fig:MultSpectraKPhi} (left), while the comparison to the EPOS v3.107 model \cite{L11} shows only a qualitative agreement with the data. The right plot in Fig.\ref{fig:MultSpectraLambda} presents the $\langle p_{T} \rangle$ as a function of the system size. Here the Blast-Wave model is again in a remarkable agreement with the measured data and the EPOS v3.107 (with UrQMD \cite{L25, L26}), which includes hadronic interactions in the final state, approximates the evolution of the data points fairly well. From this, it can be inferred that the very short lived $\Lambda(1520)$ resonance originates from the same thermodynamic source as $\pi, K$ and $p$ and is likely to have some additional interactions in the final state of the system evolution.

\begin{figure}[h!]
     \centering
     \begin{subfigure}[l]{0.59\textwidth}
         \centering
        \includegraphics[width=\textwidth]{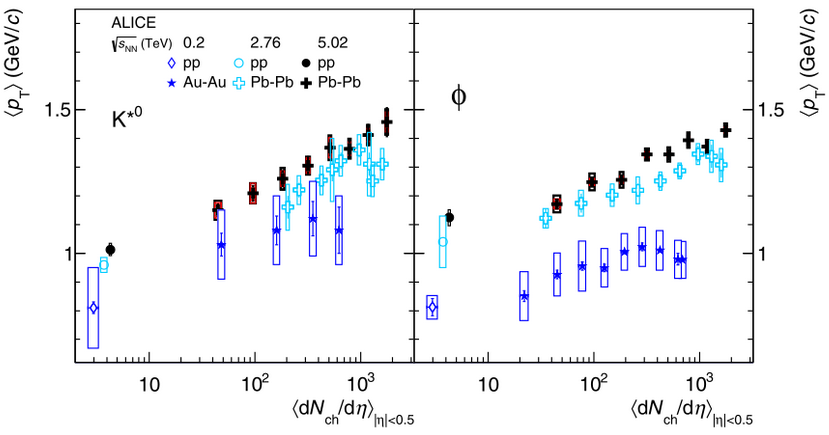}
    \end{subfigure}
    \hfill
    \begin{subfigure}[r]{0.33\textwidth}
         \centering
        \includegraphics[width=\textwidth]{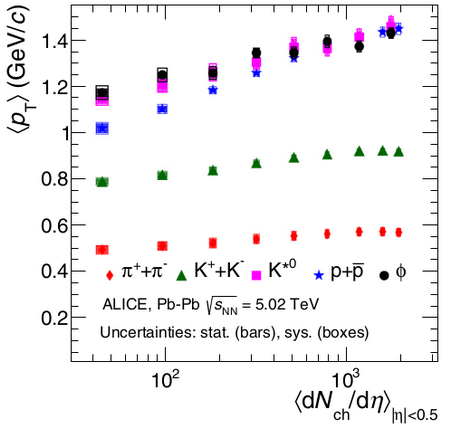}
    \end{subfigure}
    \caption{The left panel shows $\langle p_{T} \rangle$ as a function of the charged particle multiplicity $\langle dN_{ch}/d\eta \rangle$ for $K^{*}(892)^{0}$ and $\phi(1020)$ at $\sqrt{s_{NN}} = 5.02$ $\mathrm{TeV}$ compared to values from Pb–Pb and pp collisions at $\sqrt{s_{NN}} = 2.76$ $\mathrm{TeV}$ \cite{K18, K21}, and Au–Au and pp collisions at $\sqrt{s_{NN}} = 200$ $\mathrm{GeV}$ \cite{K9, K10, K12, K13} \cite{PhiK}. The right panel compares the $\langle p_{T} \rangle$ as a function of $\langle dN_{ch}/d\eta \rangle$ for $K^{*}(892)^{0}$ and $\phi(1020)$ at $\sqrt{s_{NN}} = 5.02$ $\mathrm{TeV}$ to $\pi^{\pm}, K^{\pm}$ and $p(\Bar{p})$ \cite{K27} \cite{PhiK}.}
    \label{fig:MeanpTKaPhi}
\end{figure}
The $\langle p_{T} \rangle$ as a function of $\langle dN_{ch}/d\eta \rangle$ for $K^{*}(892)^{0}$ and $\phi(1020)$ increases towards larger systems. This can be directly seen from the growing $\langle p_{T} \rangle$ as larger system sizes are approached as shown in Fig.\ref{fig:MeanpTKaPhi} (left plot)  and from the fact that the pp results, in the same plot, show generally smaller $\langle p_{T} \rangle$ compared to the heavy-ions. There is also a subtle increase of $\langle p_{T} \rangle$ with $\sqrt{s_{NN}}$. The $\langle p_{T} \rangle$ at a given $\langle dN_{ch}/d\eta \rangle$ is of the same order from $K^{*}(892)^{0}$ to $\phi(1020)$, due to their similarity in terms of their mass. Fig.\ref{fig:MeanpTKaPhi} (right) supports this observation as the increase of $\langle p_{T} \rangle$ with the $\langle dN_{ch}/d\eta \rangle$ is clearly steeper for heavier particles. This suggests growing contributions of radial flow with increasing $\langle dN_{ch}/d\eta \rangle$ and triggers the question of production time of these resonances as there is a breaking of mass ordering at low charged particle multiplicities.

\subsection{Ratio to stable particles}

\begin{figure}[h!]
     \centering
     \begin{subfigure}[l]{0.5\textwidth}
         \centering
        \includegraphics[width=\textwidth]{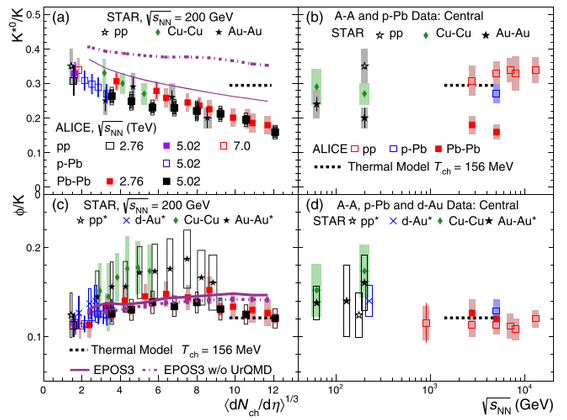}
    \end{subfigure}
    \hfill
    \begin{subfigure}[r]{0.35\textwidth}
         \centering
        \includegraphics[width=\textwidth]{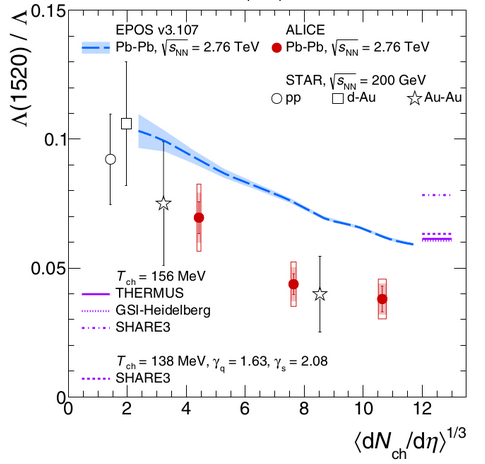}
    \end{subfigure}
    \caption{Model comparisons for $K^{*}(892)^{0}/K$ and $\phi(1020)/K$ as a function of $\langle dN_{ch}/d\eta \rangle^{1/3}$ (panels (a) and (c)) and $\sqrt{s_{NN}}$ (panels (b) and (d) for several collision systems and energies \cite{PhiK, K9, K10, K12, K13, K14, K18, K21, K49, K50, K51, K70}; Grand-canonical thermal model from \cite{K39} and the EPOS3 model predictions from \cite{K20}. Model comparisons from GSI-Heidelberg \cite{L4}, THERMUS \cite{L29}, SHARE3 \cite{L30} (parameters from fit to stable particles \cite{L31} and non-equilibrium configuration \cite{L32}) and EPOS3 to the data from RHIC in Au-Au, d-Au and pp collisions at $\sqrt{s_{NN}} = 200$ $\mathrm{GeV}$ \cite{L13, L14} for $\Lambda(1520)/\Lambda$ as a function of $\langle dN_{ch}/d\eta \rangle^{1/3}$ (right plot) \cite{Lambda}.}
    \label{fig:Ratio}
\end{figure}
From the left set of plots in Fig.\ref{fig:Ratio} it can be seen that all ratios are larger for pp-data with respect to the larger heavy-ion systems. The $K^{*}(892)^{0}/K$ ratio as a function of system size (Fig.\ref{fig:Ratio} (a)) shows a clear suppression for larger systems whereas the $\phi(1020)/K$ (Fig.\ref{fig:Ratio} (c)) is approximately constant. Thermal models clearly underestimate the $K^{*}(892)^{0}/K$ but approximate the $\phi(1020)/K$, as well as the energy dependence of both ratios. The two different tunes of EPOS3, one with a hadronic afterburner and the other one without, underline the relevance of the hadronic interactions in the final state, as only the model with UrQMD can describe the constant $\phi(1020)/K$ and the suppressed $K^{*}(892)^{0}/K$ ratio. These observations from the $K^{*}(892)^{0}$ and $\phi(1020)$ results \cite{PhiK} are consistent with those from the $\Lambda(1520)$ \cite{Lambda}. On the right plot in Fig.\ref{fig:Ratio}, the $\Lambda(1520)/\Lambda$ ratio shows a similar suppression as the system size increases. This cannot be described by thermal models alone but by EPOS3 configured with UrQMD. Comparing the lifetimes of the analyzed resonances, there is a clear suppression of resonance ratios towards larger system size for the very short lived $K^{*}(892)^{0}$ and $\Lambda(1520)$ but not for the $\phi(1020)$. This supports the assumption that the $\phi(1020)$ might be rather unaffected by re-scattering and re-generation effect in the hadronic-gas phase.

\subsection{Conclusion}
There are multiple indications of the importance of hadronic interactions in the final state of the system produced in heavy-ion collisions. Even though there are hints of a subtle collision energy dependence, the overall resonance production is clearly driven by the event multiplicity with growing contributions of radial flow towards larger system sizes. The ratios of very short lived resonances are found to be suppressed in central collisions. These two important observations support the assumption of dominant re-scattering effects for short-lived resonances at low momenta. At the current stage, all models fail qualitatively or quantitatively to describe the hadronic resonances from proton-proton and/or heavy-ion collisions over the full $p_{T}$-range. Further developments are required to better understand the evolution of the QGP and the hadronic interactions that are likely to occur at the very late stages of the evolution of the system produced in heavy-ion collisions.

\end{document}